\documentclass[twocolumn,showpacs,preprintnumbers,amsmath,amssymb,superscriptaddress, bibnotes]{revtex4}

\usepackage{graphicx}
\usepackage{dcolumn}
\usepackage{bm}
\usepackage{xspace}

\newcommand{\EF}{$E_\mathrm{F}$\xspace}
\newcommand{\Uf}{$\mathrm{U}~5f$\xspace}
\newcommand{\Cd}{$\mathrm{Co}~3d$\xspace}
\newcommand{\hn}[1]{$h\nu= #1~\mathrm{eV}$\xspace}

\newcommand{\UG}{$\mathrm{UGe}_2$\xspace}
\newcommand{\UCG}{$\mathrm{UCoGe}$\xspace}
\newcommand{\URG}{$\mathrm{URhGe}$\xspace}

\newcommand{\etal}{\textit{et al.}\xspace}

\begin{document}
\draft
\preprint{HEP/123-qed}

\title{Electronic structures of ferromagnetic superconductors \UG and \UCG studied by angle-resolved photoelectron spectroscopy}

\author{Shin-ichi~Fujimori}
\affiliation{Condensed Matter Science Division, Japan Atomic Energy Agency, Sayo, Hyogo 679-5148, Japan}

\author{Takuo~Ohkochi}
\altaffiliation[Present address: ]{Japan Synchrotron Radiation Research Institute/SPring-8, Sayo, Hyogo 679-5198, Japan}
\affiliation{Condensed Matter Science Division, Japan Atomic Energy Agency, Sayo, Hyogo 679-5148, Japan}

\author{Ikuto~Kawasaki}
\altaffiliation[Present address: ]{Graduate School of Material Science, University of Hyogo, Kamigori, Hyogo 678-1297}
\affiliation{Condensed Matter Science Division, Japan Atomic Energy Agency, Sayo, Hyogo 679-5148, Japan}

\author{Akira~Yasui}
\altaffiliation[Present address: ]{Japan Synchrotron Radiation Research Institute/SPring-8, Sayo, Hyogo 679-5198, Japan}
\affiliation{Condensed Matter Science Division, Japan Atomic Energy Agency, Sayo, Hyogo 679-5148, Japan}

\author{Yukiharu~Takeda}
\affiliation{Condensed Matter Science Division, Japan Atomic Energy Agency, Sayo, Hyogo 679-5148, Japan}

\author{Tetsuo~Okane}
\affiliation{Condensed Matter Science Division, Japan Atomic Energy Agency, Sayo, Hyogo 679-5148, Japan}

\author{Yuji~Saitoh}
\affiliation{Condensed Matter Science Division, Japan Atomic Energy Agency, Sayo, Hyogo 679-5148, Japan}

\author{Atsushi~Fujimori}
\affiliation{Condensed Matter Science Division, Japan Atomic Energy Agency, Sayo, Hyogo 679-5148, Japan}
\affiliation{Department of Physics, University of Tokyo, Hongo, Tokyo 113-0033, Japan}

\author{Hiroshi~Yamagami}
\affiliation{Condensed Matter Science Division, Japan Atomic Energy Agency, Sayo, Hyogo 679-5148, Japan}
\affiliation{Department of Physics, Faculty of Science, Kyoto Sangyo University, Kyoto 603-8555, Japan}

\author{Yoshinori~Haga}
\affiliation{Advanced Science Research Center, Japan Atomic Energy Agency, Tokai, Ibaraki 319-1195, Japan}

\author{Etsuji~Yamamoto}
\affiliation{Advanced Science Research Center, Japan Atomic Energy Agency, Tokai, Ibaraki 319-1195, Japan}

\author{Yoshichika~\=Onuki}
\affiliation{Advanced Science Research Center, Japan Atomic Energy Agency, Tokai, Ibaraki 319-1195, Japan}
\affiliation{Faculty of Science, University of the Ryukyus, Nishihara, Okinawa 903-0213, Japan}

\date{\today}

\begin{abstract}
The electronic structures of the ferromagnetic superconductors \UG and \UCG in the paramagnetic phase were studied by angle-resolved photoelectron spectroscopy using soft X-rays (\hn{400-500}).
The quasi-particle bands with large contributions from \Uf states were observed in the vicinity of \EF, suggesting that the \Uf electrons of these compounds have an itinerant character.
Their overall band structures were explained by the band-structure calculations treating all the \Uf electrons as being itinerant.
Meanwhile, the states in the vicinity of \EF show considerable deviations from the results of band-structure calculations, suggesting that the shapes of Fermi surface of these compounds are qualitatively different from the calculations, possibly caused by electron correlation effect in the complicated band structures of the low-symmetry crystals.
Strong hybridization between \Uf and $\mathrm{Co}~3d$ states in \UCG were found by the $\mathrm{Co}~2p-3d$ resonant photoemission experiment, suggesting that $\mathrm{Co}~3d$ states have finite contributions to the magnetic, transport, and superconducting properties.
\end{abstract}

\maketitle
\narrowtext
\section{INTRODUCTION}
The coexistence of ferromagnetic ordering and superconductivity is one of the most intriguing phenomena in uranium compounds\cite{HF_SC}. 
Four inter-metallic uranium compounds, \UG, $\mathrm{UIr}$, \URG, and \UCG have been known as such ferromagnetic superconductors \cite{AokiReview}.
Recently, unconventional critical behaviors of magnetization were reported in \UG and \URG, suggesting that ferromagnetism and superconductivity are closely related in these compounds\cite{Tateiwa}. 
Although the nature of magnetism and superconductivity of these compounds has been well studied so far, their electronic structures have not been well understood due to the lack of experimental electronic structure studies.
In the previous studies, we have studied the electronic structures of $\mathrm{UIr}$ and \URG by angle-resolved photoelectron spectroscopy\cite{UIr_ARPES,URhGe_ARPES}.
It was found that the \Uf electrons of these compounds have an itinerant character.

In the present study, we have further studied the electronic structures of ferromagnetic uranium superconductors \UG and \UCG by angle-resolved photoelectron spectroscopy (ARPES) to understand their electronic structures.
Both of \UG and \UCG have the orthorhombic-type crystal structures.
\UG has the $\mathrm{ZrGa}_2$-type crystal structure of the space group $Cmmm$ with a large lattice constant along the $b$-axis\cite{UGe2_crystal}.
It undergoes a ferromagnetic transition at $T_{\rm Curie}=52~\mathrm{K}$ and a superconducting transition at $T_{\rm SC}=1.2~\mathrm{K}$ under the pressure of $P_{\rm C}=1.2~\mathrm{GPa}$ \cite{UGe2_SC}.
The itinerant nature of the \Uf electrons in this compound has been argued based on the small ordered moment in the ferromagnetic phase\cite{HF_SC} and the result of the de Haas-van Alphen study\cite{UGe2_dHvA}.
Meanwhile, the positron annihilation study of \UG suggests that the fully localized \Uf model can explain their experimental result\cite{UGe2_positron}.
Furthermore, the dual nature of the \Uf electrons in \UG has been proposed for this compound from the macroscopic study made in wide temperature and magnetic-field ranges\cite{UGe2_dual}.
The X-ray photoelectron spectroscopy (XPS) study of \UG also suggested that \Uf electrons have the dual (itinerant and localized) character\cite{UGe2_XPS}.
Therefore, the nature of \Uf state in \UG still has been a controversial issue.

\UCG has the $\mathrm{TiNiSi}$-type crystal structure of the space group $Pnma$.
It undergoes a ferromagnetic transition at $T_{\rm Curie}=3~\mathrm{K}$, and a superconducting transition at $T_{\rm SC}=0.8~\mathrm{K}$ at an ambient pressure\cite{UCoGe}.
The magnitude of the magnetic moment is $\mu_{\rm ord}=0.03~\mu_{\rm B}$, suggesting that \UCG is an itinerant weak ferromagnet. 
Meanwhile, there are only few electronic structure studies on this compound.
Samsel-Czeka\l a \etal reported the XPS study of \UCG \cite{UCoGe_XPS}.
They compared their valence-band XPS spectrum with the band-structure calculation, and the calculated density of states fairly well explain the spectrum. 
Aoki \etal measured the Shubnikov-de Haas oscillations of \UCG, and suggested that \UCG is a low carrier system\cite{UCoGe_SdH}.
Meanwhile, its detailed electronic structure has not been well understood, and it is desired to observe its electronic structure by ARPES.

\begin{figure*}
\includegraphics[scale=0.42]{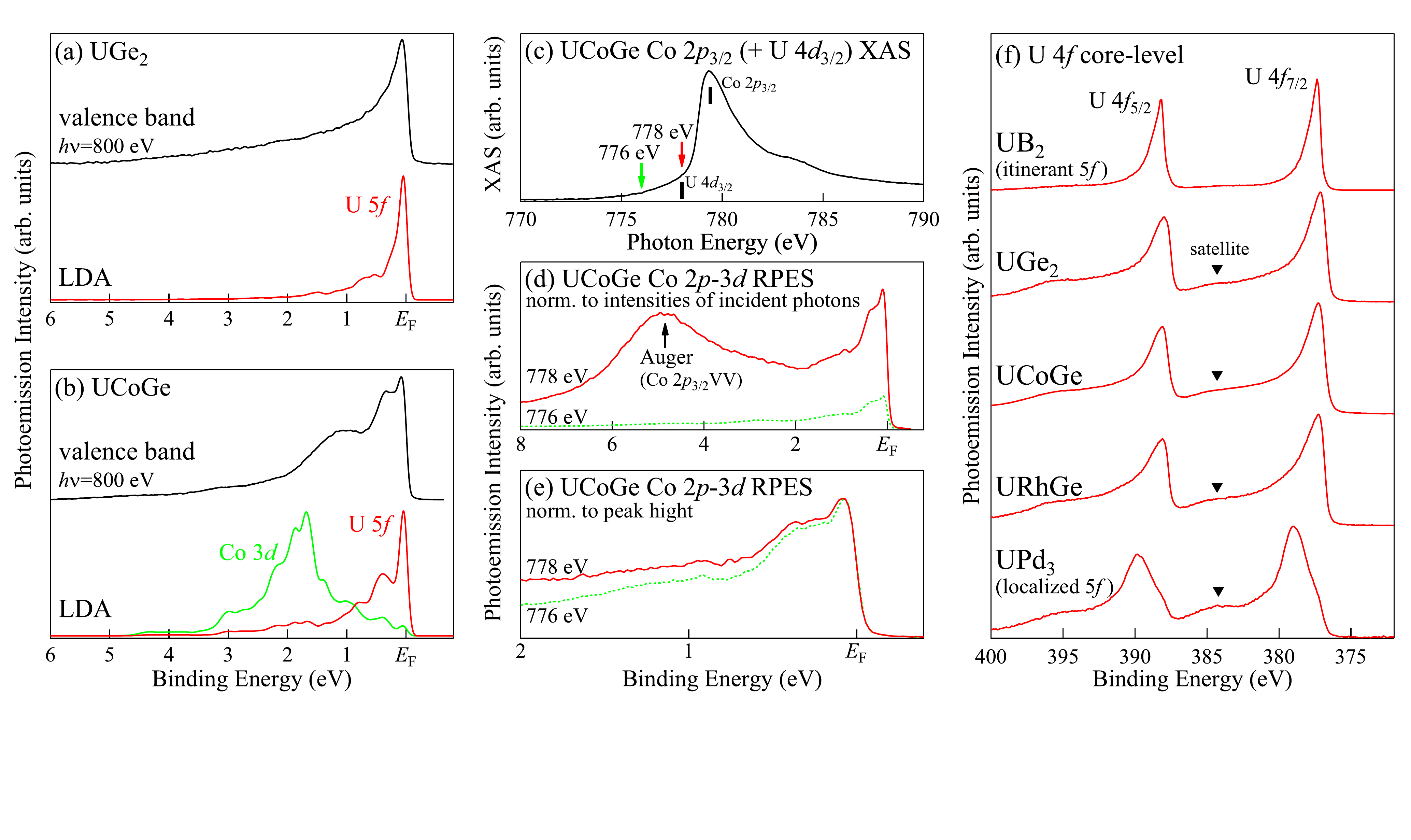}
\caption{(Online color)
Angle-integrated photoemission spectra of \UG and \UCG.
(a) Valence-band spectrum of \UG measured at \hn{800}, and the calculated \Uf partial density of states.
(b) Valence-band spectrum of \UCG measured at \hn{800}, and the calculated \Uf and \Cd partial density of states.
(c) $\mathrm{Co}~2p_{3/2}$ and $\mathrm{U}~4d_{3/2}$ X-ray absorption spectrum of \UCG.
Positions of $\mathrm{Co}~2p_{3/2}$ and $\mathrm{U}~4d_{3/2}$ absorption edges are indicated by solid lines.
(d) Valence-band spectra of \UCG measured at \hn{776} (off-resonance) and $778~\mathrm{eV}$ (on-resonance).
The spectra have been normalized to the intensities of incident photons.
The on-resonance spectrum is off the peak top since the Auger signals become significant with further increase of photon energies. 
(e) The same spectra normalized to the peak height.
(f) $\mathrm{U}~4f$ core-level spectra of \UG, \UCG, and \URG together with those of the typical itinerant \Uf compound $\mathrm{UB}_2$ and the localized compound $\mathrm{UPd}_3$.
}
\label{AIPES}
\end{figure*}

\section{EXPERIMENTAL PROCEDURES}
Photoemission experiments were performed at the soft X-ray beamline BL23SU of SPring-8 \cite{BL23SU,BL23SU2}.
The overall energy resolution in angle-integrated photoemission (AIPES) experiments at \hn{800} was about $110~\mathrm{meV}$, and that in ARPES experiments at \hn{400-500} was $80-90~\mathrm{meV}$, depending on the photon energies.
The position of \EF was carefully determined by measurements of the vapor-deposited gold film.
Clean sample surfaces were obtained by cleaving the samples {\it in situ} with the surface parallel to the $ac$ plane for \UG and the $ab$ plane for \UCG under ultra-high vacuum (UHV) condition.
The vacuum during the course of measurements was typically $<3 \times 10^{-8}~\mathrm{Pa}$, and the sample surfaces were stable for the duration of measurements (1-2 days) since no significant changes had been observed in ARPES spectra during the periods.
The positions of ARPES cuts were determined by assuming a free-electron final state with an inner potential of $V_{0}=12~\mathrm{eV}$ in both compounds.

\section{RESULTS}
\subsection{Angle-integrated photoemission spectra of \UG and \UCG}
First, we present the AIPES spectra of \UG and \UCG in the paramagnetic phases.
Figure~\ref{AIPES} (a) and (b) show the valence-band spectra of \UG and \UCG.
The photon energy was \hn{800}, and the sample temperatures were $120~\mathrm{K}$ (\UG) and $20~\mathrm{K}$ (\UCG), and both samples were in the paramagnetic phase.
These spectra are identical to those in Ref.\cite{Ucore}.
In this photon energy range, the contributions from \Uf and \Cd states are dominant, and those from $\mathrm{Ge}~s$ and $p$ states are two or three orders of magnitude smaller than the \Uf and \Cd states \cite{Atomic}.
The spectrum of \UG is very similar to the $5d-5f$ resonant photoemission spectrum \cite{UHAXPES}, suggesting that the \Uf contribution is dominant in this photon energy range.
Furthermore, these photon energies are high enough to probe the bulk electronic structures of uranium based compounds \cite{UN_ARPES}.

In Figs.~\ref{AIPES} (a) and (b), the calculated partial density of states (DOS) are also shown at the bottom.
In the calculation, relativistic linear augmented-plane-wave (RLAPW) band-structure calculations\cite{Yamagami} within the local
density approximation (LDA)\cite{LDA} were performed for \UG and \UCG treating all \Uf electrons as being itinerant.
These calculated DOS have been broadened with the experimental energy resolution.
Here, it should be noted that the present experimental spectra are sharper than those of the previous studies\cite{UGe2_XPS, UCoGe_XPS}.
This might be due to the much better energy resolution as well as the oxygen-free sample surface in the present study as discussed below.

The spectrum of \UG shows an asymmetric line shape, having a sharp peak structure just below \EF.
The \Uf states have a large contribution to \EF, suggesting that \Uf electrons have an itinerant character in this compound.
The calculated DOS have very similar asymmetric line shape, and the agreement between the experiment and the calculation is fairly good.
The spectrum of \UCG also shows a sharp peak structure just below \EF.
In addition to this peak, there is a broad peak at around $E_{\mathrm{B}} \sim 1~\mathrm{eV}$.
Comparison with the calculated DOS suggests that this peak mainly originates from \Cd states.
An interesting point to note is that there is a shoulder structure at around $E_{\mathrm{B}} \sim 0.5~\mathrm{eV}$.
Similar shoulder structure was also observed in the valence-band spectrum of \URG \cite{URhGe_ARPES}.
In strongly correlated electron systems, an incoherent satellite peak has been observed on the high-binding-energy side of the main peak, and this shoulder structure is very similar to it.
However, similar peak structure exists in the calculated DOS, suggesting that this peak originates not from correlation effects but also from the band structure of this compound. 
The overall spectral shape is well explained by the band-structure calculation.

To identify the \Cd contributions in the valence-band spectrum, we have carried out the $\mathrm{Co}~2p-3d$ resonant photoemission experiment.
In the resonant photoemission process, the photo-ionization cross-section of specific orbitals can be enhanced by tuning the photon energy to that of the absorption energy.
In the present study, we have utilized the $\mathrm{Co}~2p_{3/2}$ absorption edge to enhance the photoionization cross section of \Cd orbitals.

Figure~\ref{AIPES} (c) shows the $\mathrm{Co}~2p_{3/2}$ X-ray absorption spectrum (XAS) of \UCG.
In this energy region, the $\mathrm{U}~4d_{3/2}$ absorption edge also exists, but its magnitude is much smaller than that of the $\mathrm{Co}~2p_{3/2}$ absorption edge\cite{UCoAl_MCD}.
Furthermore, it has been pointed out that the enhancement in the $\mathrm{U}~4d-5f$ resonance is negligible in the $\mathrm{U}~4d$ absorption edge\cite{Allen_RPES}, and the enhancement can be attributed to the enhancement of the \Cd cross section.
The XAS spectrum has a peak at \hn{779.3}.
By tuning the photon energy to this energy, the cross section from \Cd state should be enhanced.

Figure~\ref{AIPES} (d) shows the angle integrated photoemission spectra measured at \hn{776} and $778~\mathrm{eV}$.
The spectra have been normalized to the intensities of incident photons.
It is shown that the spectrum measured at \hn{778} show a strong enhancement in the intensity of the photoemission spectral function.
Meanwhile, the spectrum has a large peak at around $E_{\mathrm{B}} \sim 5$~eV, which does not exists in the spectrum measured at \hn{776}.
This is the contribution from the $\mathrm{Co}~2p_{3/2} VV$ Auger peak since its peak position moves toward high binding energies as the photon energy is increased.
This Auger signal is too enhanced by further increases of photon energies that the shape of the valence band spectrum cannot be recognized.
The situation can be understood from the case of the $\mathrm{Co}~2p-3d$ resonant photoemission study of $\mathrm{CoSb}_3$ \cite{Co_RPES}.
Therefore, we use the spectra measured at \hn{776} and $778~\mathrm{eV}$ as off-resonance and on-resonance spectra respectively although \hn{778} is off the peak top .

Figure~\ref{AIPES} (e) shows the comparison of the valence-band spectra measured at \hn{776} and $778~\mathrm{eV}$, normalized to the peak height at \EF.
Here, an important point to note is that both spectra have very similar spectral line shapes including the states just below \EF, which have large contributions from the \Uf states.
This means that the partial \Uf and \Cd DOS have very similar shapes, suggesting that they are strongly hybridized in \UCG.

Figure~\ref{AIPES} (f) shows the core-level spectra of \UG, \UCG, and \URG together with the typical itinerant compound $\mathrm{UB}_2$ and the localized compounds $\mathrm{UPd}_3$.
Those spectra are identical to those in Ref.\cite{Ucore}.
Here, it should be noted that the spectrum of $\mathrm{UPd}_3$ is very similar to that of HAXPES study \cite{UHAXPES}.
The spectra show a spin-orbit splitting corresponding to $\mathrm{U}~4f_{7/2}$ and $\mathrm{U}~4f_{5/2}$ components, and each of them consists of the dominant main line and the broad satellite located on the high-binding-energy side.
The core-level spectra of \UG, \UCG, and \URG have asymmetric line shapes, and are very similar to that of the itinerant compound $\mathrm{UB}_2$ rather than that of $\mathrm{UPd}_3$ .
This suggests that \Uf electrons in \UG, \UCG, and \URG essentially have itinerant character.
Meanwhile, there is a weak but finite satellite structure on the high-binding-energy side, suggesting that a weak correlation effect also exists in these compounds.

Here, we compare the present results with previous photoemission studies.
The XPS studies of \UG \cite{UGe2_XPS} and \UCG \cite{UCoGe_XPS} have suggested that their valence-band spectra are much broader than those predicted by the band-structure calculations.
Moreover, their $\mathrm{U}~4f$ core-level spectra showed two satellite structures on the high-binding-energy side of the main lines.
They claimed that the electron correlation effect is the origin of these broadening and the satellite structure of photoemission spectral line shapes.
Meanwhile, their valence-band and $\mathrm{U}~4f$ core-level spectra were much broader than those of the present study even the difference of the energy resolutions is taken into account.
In particular, the satellite peaks designated as component II in \UG ($E_{\mathrm{B}} = 380.7~\mathrm{eV}$)\cite{UGe2_XPS} and sat. 2 in \UCG ($E_{\mathrm{B}} = 380.2~\mathrm{eV}$)\cite{UCoGe_XPS} in their $\mathrm{U}~4f$ core-level spectra are absent in the present study.
Since their peak positions were very close to those of $\mathrm{UO}_2$ ($E_{\mathrm{B}} = 380.2~\mathrm{eV}$), these components presumably originated from the oxidized components in their sample surfaces, which broaden their valence-band spectra too.

\subsection{Band structure and Fermi surfaces of \UG in the paramagnetic phase}
\begin{figure*}[t]
\includegraphics[scale=0.45]{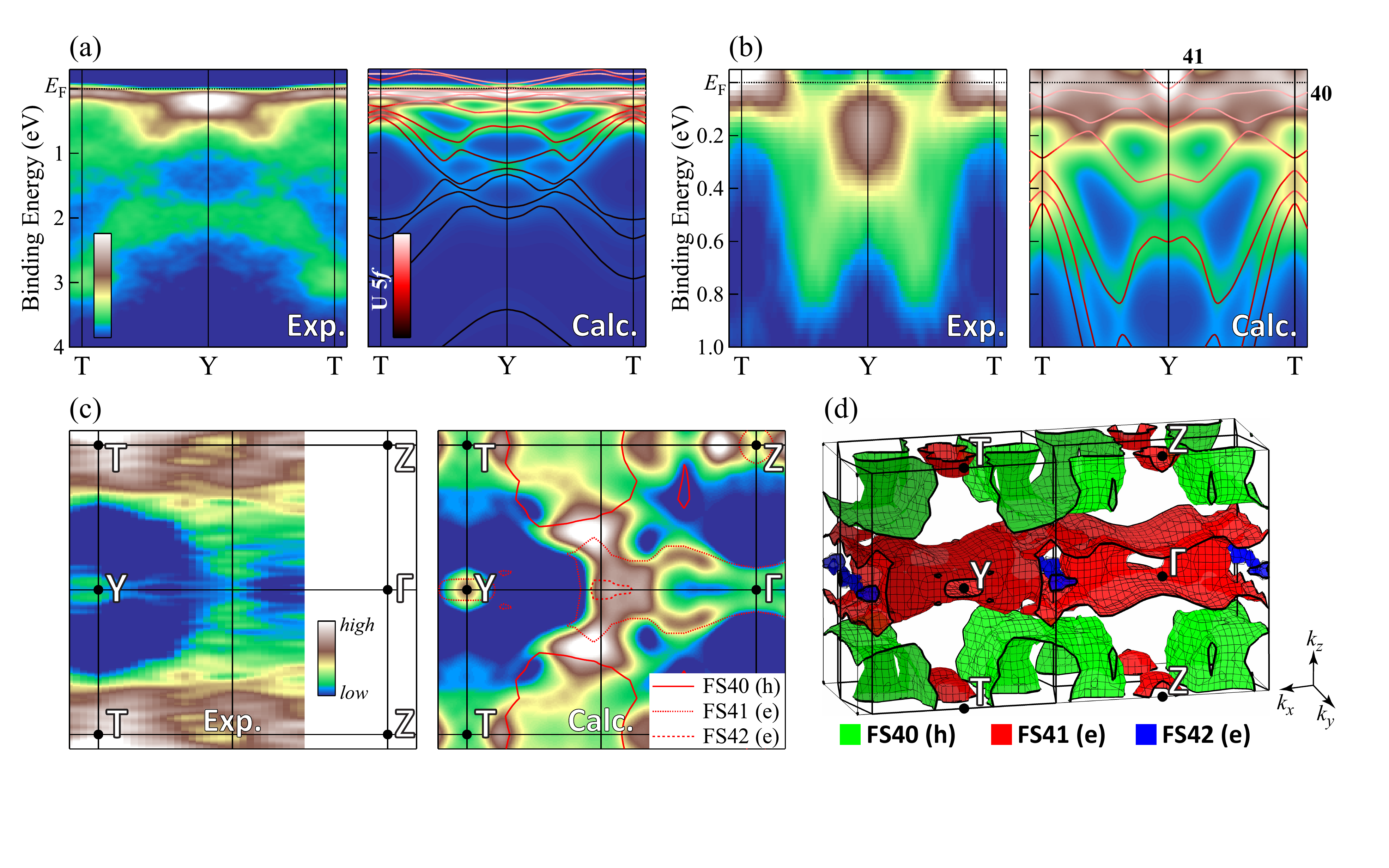}
\caption{(Online color)
ARPES spectra and Fermi surface mapping of \UG in the paramagnetic phase.
(a) ARPES spectra measured along the $\mathrm{T}-\mathrm{Y}-\mathrm{T}$ high-symmetry line (left), and the calculated energy bands and the simulation of ARPES spectra based on band-structure calculation (right).
(b) Blowup of experimental ARPES spectra (left), and the corresponding energy bands and the simulation based on band-structure calculation (right).
(c) Fermi surface mapping of ARPES spectra (left) and the simulation of Fermi surface mapping based on the band-structure calculation (right).
The lines represent the calculated Fermi surface.
(d) Brillouin zone and calculated Fermi surfaces.
}
\label{UGe2_ARPES_FS}
\end{figure*}
First, we show the band structure of \UG in the paramagnetic phase.
The left panel of Fig.~\ref{UGe2_ARPES_FS} (a) shows the ARPES spectra of \UG measured along the $\mathrm{T}-\mathrm{Y}-\mathrm{T}$ high-symmetry line.
The sample temperature was kept at $120~\mathrm{K}$, and the sample was in the paramagnetic phase.
The photon energy was \hn{437}, which was chosen by measuring the photon energy dependence of ARPES spectra. 
These spectra are symmetrized relative to the high-symmetry points to eliminate the photoemission structure factor effect \cite{PSF}.
The detail is described in Refs.~\cite{UN_ARPES, URhGe_ARPES}.
Furthermore, the background contributions due to the elastically scattered photoelectrons have been subtracted.
The detail of this procedure is described in Appendix A.

In the ARPES spectra, energy dispersions were observed.
On the high-binding-energy side ($E_{\mathrm{B}} \gtrsim 0.5$~eV ), there exist strongly dispersive bands.
Those are bands with large contributions from $\mathrm{Ge}~s$ and $p$ states.
Meanwhile, there exist weakly dispersive bands in the vicinity of \EF.
These are contributions mainly from \Uf quasi-particle bands.
The right panel of Fig.~\ref{UGe2_ARPES_FS} (a) shows the calculated band structure and the simulation of ARPES spectra based on the band-structure calculation.
Solid lines represent the band dispersions and the image does the simulation.
The color coding of the bands is the projection of the contributions from \Uf states.
In the simulation, the following effects were taken into account: (i) the broadening along the $k_\perp$ direction due to the finite escape depth of photoelectrons, (ii) the lifetime broadening of the photohole, (iii) the photoemission cross sections of orbitals, and (iv) the energy resolution and the angular resolution of the electron analyzer.
The details are described in Ref.~\cite{UN_ARPES}.
Some similarities are recognized between the ARPES spectra and the calculation.
On the high-binding-energy side, there are broad and strongly dispersive features in both the experiment and the calculation.
In particular, the inverted parabolic dispersions centered at the $\mathrm{Y}$ point at $E_{\mathrm{B}}=1-3.5~\mathrm{eV}$ agree very well between the experiment and the calculation. 
In addition, the parabolic dispersions with their bottoms at around $E_{\mathrm{B}} \sim 2~\mathrm{eV}$ centered at the $\mathrm{Y}$ point are recognized in both in the experiment and the calculation.
Meanwhile, in the vicinity of \EF, there exist less dispersive bands both in the experiment and the calculation, and their agreements are less clear.

To understand the details of the band-structure in the vicinity of \EF, the blowup of the experimental ARPES spectra is shown in the left panel of Fig.~\ref{UGe2_ARPES_FS} (b).
These spectra have been divided by the Fermi-Dirac function broadened by the instrumental energy resolution to reveal the states near \EF more clearly.
There exist narrow dispersive bands in the vicinity of \EF, suggesting that \Uf quasi-particle bands form Fermi surfaces in this compound.
They have a hole-type dispersion around the $\mathrm{T}$ point, and they hybridize with the inverted parabolic bands at the $\mathrm{Y}$ point in $E_{\mathrm{F}} \sim 0.2~\mathrm{eV}$.
Meanwhile, in the band-structure calculation shown in the right panel of Fig.~\ref{UGe2_ARPES_FS} (b), there exist a number of narrow bands with large contributions from \Uf states in the vicinity of \EF.
In the band-structure calculation, band 41 form an electron-type Fermi surface around the $\mathrm{Y}$ point.
It is shown that the band-structure calculation predicts very complicated features especially in the vicinity of \EF, and the simulation suggests that it is very difficult to distinguish each bands with the present experimental energy and momentum resolutions.
Although comparison for each band is very difficult, the comparison between the experimental spectra and the simulation suggests that there are some common features in both the experiment and the calculation.
On the other hand, the global agreement between the experiment and the calculation is far from satisfactory, suggesting that the electronic structure in the vicinity of \EF is very different from that of the band-structure calculation.

\begin{figure*}
\includegraphics[scale=0.4]{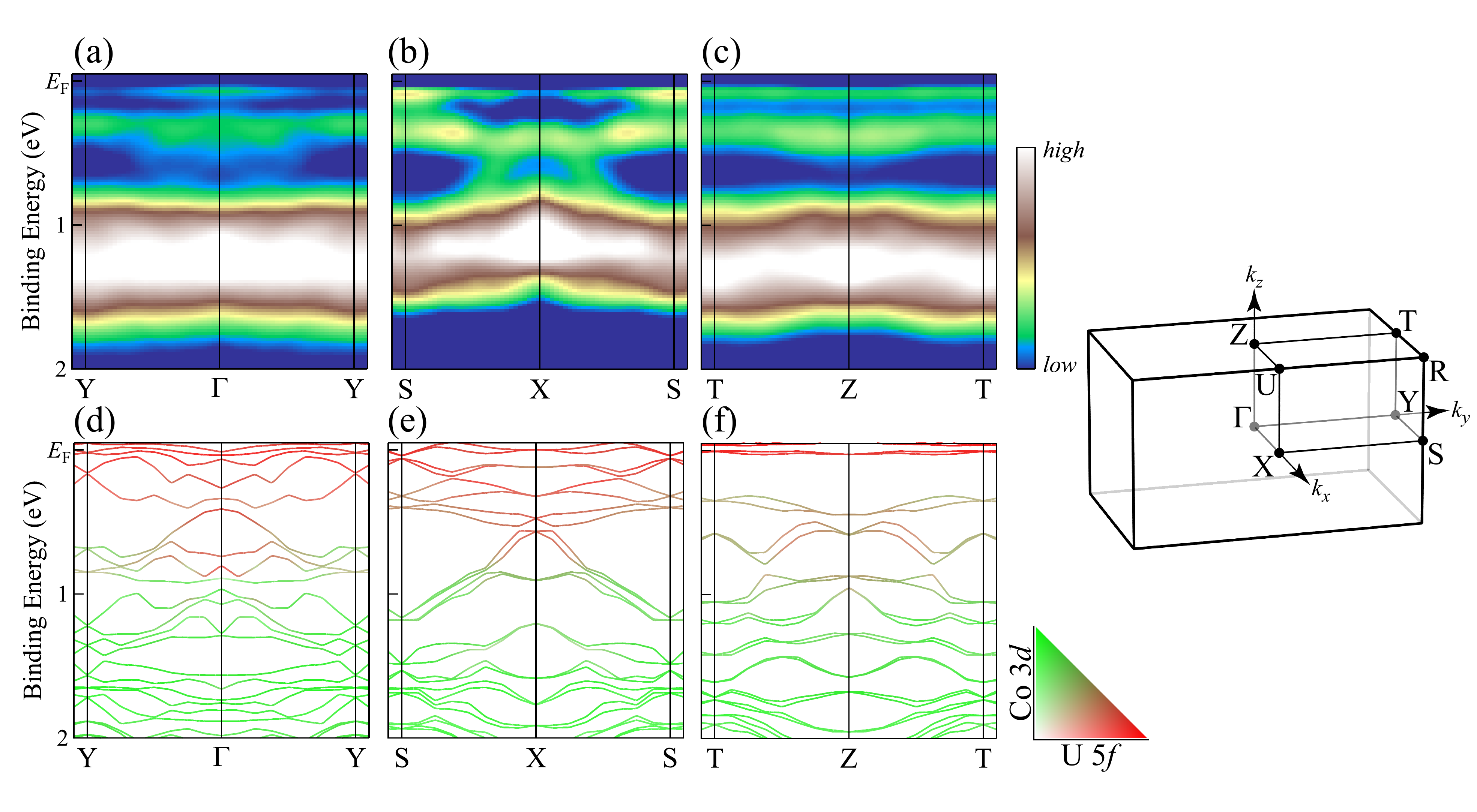}
\caption{(Online color)
Symmetrized ARPES spectra and the results of band-structure calculation of \UCG in the paramagnetic phase.
(a-c) Symmetrized ARPES spectra.
(d-f) Results of the band-structure calculation.
The color coding of bands is the projection of the contributions from \Uf states and \Cd states, respectively.
}
\label{UCoGe_ARPES1}
\end{figure*}
To reveal the nature of the Fermi surface of \UG, we have performed Fermi surface mapping as a function of momenta parallel to the sample surface.
The left panel of Fig.~\ref{UGe2_ARPES_FS} (c) shows an intensity map of ARPES spectra measured at $h\nu=437~\mathrm{eV}$.
This photon energy was chosen by measuring the photon energy dependence of ARPES spectra.
Photoemission intensities within $E_{\mathrm{F}} \pm 50~\mathrm{meV}$ of ARPES spectra were integrated and mapped as a function of $k_x$ and $k_z$.
Although a round-shaped high-intensity part is recognized around the $\mathrm{Y}$ point, it should be noted that the high-intensity part in this image does not always correspond to the position of $k_{\mathrm{F}}$ since bands in the vicinity of \EF might contribute the image even they do not cross \EF.

To understand the validity of the band-structure calculation, we compare this image with the simulation based on the band-structure calculation.
The right panel of Fig.~\ref{UGe2_ARPES_FS} (c) and Fig.~\ref{UGe2_ARPES_FS} (d) show the simulation of Fermi surface mapping based on the band-structure calculation and three-dimensional calculated Fermi surface respectively.
It should be noted that the calculated Fermi surface has somewhat different shapes from the result of the previous calculation \cite{UGe2_dHvA_Settai}.
The difference is mainly due to the accuracy of the numerical calculations.
For example, the present calculation uses much finer mesh in the momentum space, and it predicts much detailed shape of the Fermi surface than in the previous one.
Band 40 forms a hole Fermi surface which is connected along the $k_y$ direction.
Band 41 forms a large sheet-like Fermi surface along the $k_{x}-k_{y}$ directions, and a small spherical one around the $\mathrm{Z}$ point.
In the former one, each sheet is connected along the $k_z$ direction around the $\mathrm{Y}$ point.
Band 42 forms a small cylindrical Fermi surface along the $k_y$ direction.

Comparison between the experimental Fermi surface mapping and the simulation shown in Fig.~\ref{UGe2_ARPES_FS} (c) shows that there is a certain similarities between them.
For example, the round shaped feature around the $\mathrm{Y}$ point from the experiment has a similar corresponding one in the simulation.
Meanwhile, the simulation predicts much complicated variations of intensities, and the overall agreement is not satisfactory.
As shown in the right panels of Figs.~\ref{UGe2_ARPES_FS} (a) and (b), the calculation predicts many flat bands in the vicinity of \EF, and tiny changes in the band structures cause a drastic change of the Fermi surface topology.
Accordingly, it is expected that the shapes of the Fermi surfaces of \UG are qualitatively different from those obtained by the band-structure calculation.
Here, we note that the present ARPES spectra did not show significant changes below and above $T_{\mathrm{Currie}} =30\mathrm{K}$ (not shown) although a high energy resolution photoemission experiment of \UG \cite{UGe2_UPS} showed that the DOS within $E_{\mathrm{B}} \lesssim 25~\mathrm{meV}$ show a systematic change.

\begin{figure*}
\includegraphics[scale=0.4]{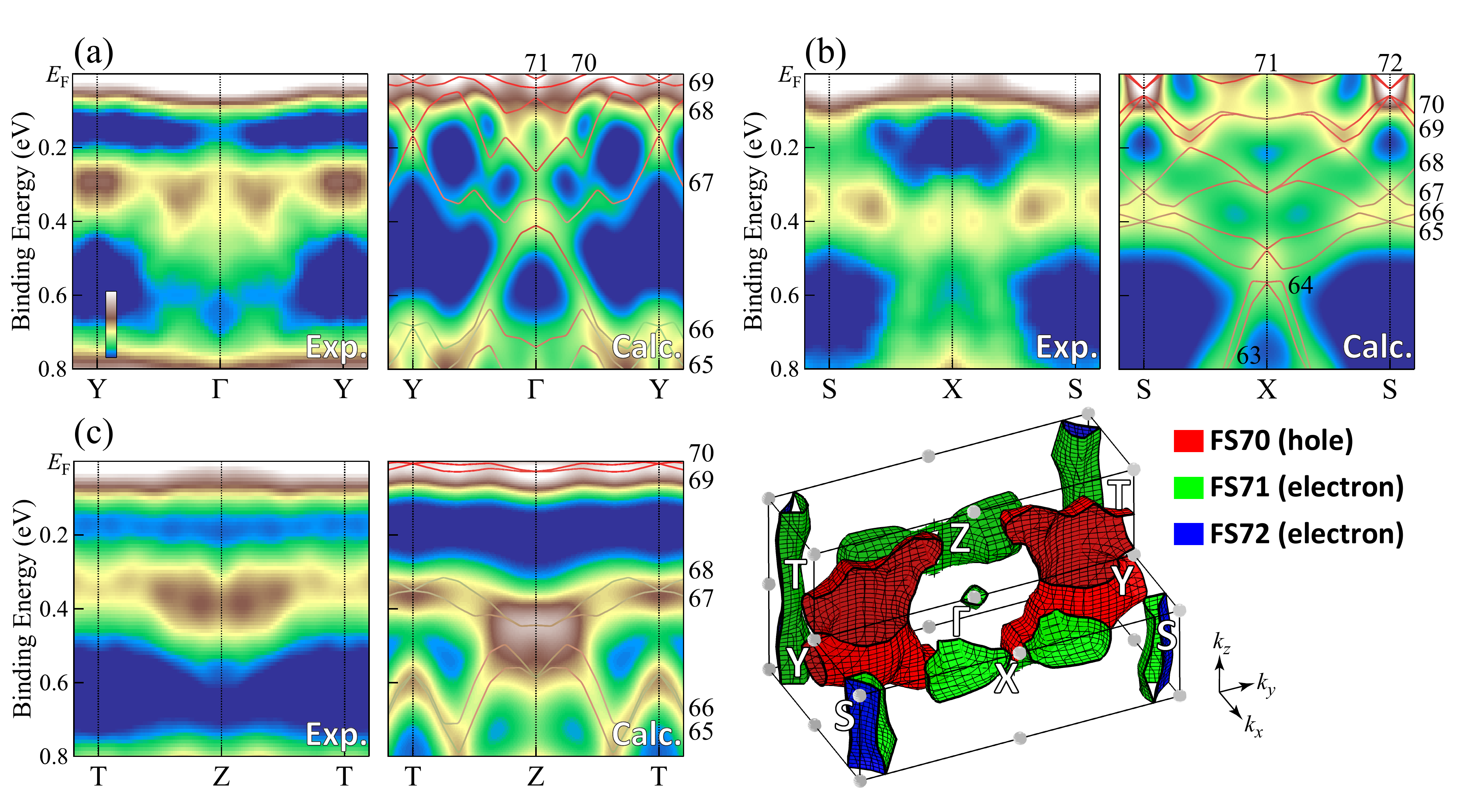}
\caption{(Online color)
Blowup of the experimental ARPES spectra of \UCG and the simulations based on band-structure calculations along the (a) $\mathrm{Y}-\mathrm{\Gamma}-\mathrm{Y}$, (b) $\mathrm{S}-\mathrm{X}-\mathrm{S}$, and (c) $\mathrm{T}-\mathrm{Z}-\mathrm{T}$ high-symmetry lines.
The calculated three-dimensional Fermi surface is also shown.
The color coding of calculated bands shown by the solid lines is the same as that in Figs.~\ref{UCoGe_ARPES1} (d)-\ref{UCoGe_ARPES1} (f).
These spectra have been divided by the Fermi-Dirac function broadened by the experimental energy resolution.
}
\label{UCoGe_ARPES2}
\end{figure*}

\subsection{Band structure of \UCG in the paramagnetic phase}
Figure~\ref{UCoGe_ARPES1} (a)-(c) show ARPES spectra of \UCG.
The sample temperature was kept at 20~K, and the sample was in the paramagnetic phase.
The photon energies were \hn{475} for the $\mathrm{Y}-\mathrm{\Gamma}-\mathrm{Y}$ and $\mathrm{S}-\mathrm{X}-\mathrm{S}$ high symmetry lines, and \hn{500} for $\mathrm{T}-\mathrm{Z}-\mathrm{T}$ high-symmetry line.
Those spectra are symmetrized relative to the high-symmetry points.
The spectra have different structures depending on the position in the Brillouin zone, suggesting that this compound has a three-dimensional electronic structure.
The spectra basically consist of three prominent features located at around $E_{\rm B} = E_{\mathrm{F}}$, $0.4~\mathrm{eV}$, and $1.2-1.4~\mathrm{eV}$.
The former two are the contributions mainly from \Uf states while the last one is mainly from \Cd states although they are strongly hybridized.
These spectral line shapes are similar to those of \URG, but the position of the $d$ bands is much closer to \EF.
This leads to a stronger $f-d$ hybridization in \UCG than in \URG. 

Figures~\ref{UCoGe_ARPES1} (d)-(f) show the calculated band structures of \UCG.
In the calculation, all \Uf electrons are treated as being itinerant.
The color coding is the projection of the contributions from \Uf states and \Cd states respectively.
The contributions from \Uf states is mostly located at $E_{\mathrm{B}} \lesssim 0.5~\mathrm{eV}$ while that from \Cd states is distributed mainly below $E_{\mathrm{B}} \gtrsim0.8~\mathrm{eV}$.
Meanwhile, \Uf and \Cd states are strongly hybridized especially in the energy region of $E_{\mathrm{B}} \lesssim 0.8~\mathrm{eV}$.
Many dispersive bands exist in the calculation, and its comparison with the experimental spectra is not straightforward although essential energy positions of the bands seem to have correspondences between the experiment and the calculation.

To see the details of the band structures in the vicinity of \EF as well as their correspondence to the band-structure calculation, the blowup of the experimental ARPES spectra and their simulations are shown in Fig.~\ref{UCoGe_ARPES2}.
The calculated three-dimensional Fermi surfaces are also shown in the figure.
The band-structure calculation predicts that bands 70-72 form Fermi surfaces in this compound.
Band 70 forms a closed hole Fermi surface around the $\mathrm{Y}$ point.
Band 71 forms relatively small electron Fermi surfaces around the $\Gamma$ and $\mathrm{X}$ points. 
Band 72 and the part of band 71 form pillar-like electron Fermi surfaces at the edges of the Brillouin zone along the $k_z$ direction. 
These pillar-like Fermi surfaces are also observed in the calculated Fermi surface of \URG.
Figure~\ref{UCoGe_ARPES2} shows the comparisons between experimental ARPES spectra and their simulations along the $\mathrm{Y}-\mathrm{\Gamma}-\mathrm{Y}$ [ Fig.~\ref{UCoGe_ARPES2} (a) ], $\mathrm{S}-\mathrm{X}-\mathrm{S}$ [ Fig.~\ref{UCoGe_ARPES2} (b) ], and $\mathrm{T}-\mathrm{Y}-\mathrm{T}$ high-symmetry lines [ Fig.~\ref{UCoGe_ARPES2} (c) ].
These spectra have been divided by the Fermi-Dirac function broadened by the instrumental energy resolution to reveal the states near \EF more clearly.
The correspondences between the ARPES spectra and the calculations are more clearly recognized.
Along the $\mathrm{Y}-\mathrm{\Gamma}-\mathrm{Y}$ high symmetry line, characteristic inverted parabolic features centered at the $\Gamma$ point correspond to the calculated features originated from bands 65-67.
Along the $\mathrm{S}-\mathrm{X}-\mathrm{S}$ high symmetry line, better correspondence is recognizable.
In particular, the features located at around $E_{\mathrm{B}}=0.2-0.8$ have very similar shapes to the calculation although their energy positions are slightly different.
The spectra along the $\mathrm{T}-\mathrm{Z}-\mathrm{T}$ high-symmetry line are much more featureless, but the high-intensity part at the $\mathrm{Z}$ point centered at around $E_{\mathrm{B}}=0.4~\mathrm{eV}$ corresponds to the one in the calculation centered at around $E_{\mathrm{B}}=0.5~\mathrm{eV}$.
Furthermore, a clear gap in the intensity at around $E_{\mathrm{B}}=0.2~\mathrm{eV}$ agrees with the calculation.
Meanwhile, experimental spectra are more featureless than the calculated ones, and the agreement becomes worse especially in states near \EF in general.

\section{DISCUSSION}
The present results reveled that \Uf electrons in \UG and \UCG form quasi-particle bands in the vicinity of \EF, suggesting that \Uf electrons have essentially itinerant character.
The results are totally different from the ARPES spectra of localized compound $\mathrm{UPd}_3$, where the \Uf electrons are located on the high-binding-energy side although they have a finite energy dispersions \cite{UPd3_ARPES,UHAXPES}.
The experimental band structures of \UG and \UCG have certain similarities to the result of the band structure calculations, but their agreements are not as good as in the cases of other uranium compounds such as itinerant compounds $\mathrm{UB}_2$ \cite{UB2_ARPES}, $\mathrm{UFeGa}_5$ \cite{UFeGa5_ARPES}, and $\mathrm{UN}$ \cite{UN_ARPES} as well as the heavy Fermion superconductor $\mathrm{UPd}_{2} \mathrm{Al}_{3}$\cite{UPd2Al3_ARPES1, UPd2Al3_ARPES2} and $\mathrm{URu}_{2} \mathrm{Si}_{2}$\cite{URu2Si2_SXARPES}, where the essential band structures as well as the topologies of Fermi surface were explained by the calculation.
In particular, the states in the vicinity of \EF are considerably different from the band-structure calculations in \UG and \UCG.
This suggests that the shapes of the Fermi surfaces might be qualitatively different from those of the band-structure calculations.
This overall agreement in the band structure and qualitative disagreement in the state in the vicinity of \EF are also recognized in the case of \URG \cite{URhGe_ARPES}.
Therefore, this seems to be a common trend of the ferromagnetic uranium superconductors.

Here, we consider possible origins of these discrepancies by taking the case of \UG.
One possible contribution is originated from the limitation in the numerical calculation of electronic structure calculations.
The band-structure calculations of these compounds predict many weakly dispersive bands especially in the vicinity of \EF, and some of them form very complicated Fermi surfaces.
This is due to the low symmetry nature of their crystal structures which removes the degeneracies of bands in these compounds.
For example, the bottom of band 41 shown in Fig.~\ref{UGe2_ARPES_FS} (b) is located at around $E_{\mathrm{B}} =10~\mathrm{meV}$, and tiny changes in the shape of energy dispersion can drastically alter the shape of Fermi surface.
In fact, the LDA calculation performed for \UG in the paramagnetic phase by Biasini and Troc \cite{UGe2_positron} gives very different shapes of Fermi surface. 
They concluded that the fully localized model can explain the result of their positron annihilation study, but this may be caused by these difficulties in the band structure calculations. 
Since there exit similar very narrow bands in the vicinity of \EF in the cases of \UCG, such discrepancies in the LDA calculations are recognized between the present calculation and the one in Ref. \cite{UCoGe_XPS}.
Therefore, the shapes of Fermi surfaces are extremely sensitive to tiny changes of the band structure, and the quantitative prediction of their electronic structures by the band structure calculation is still very difficult even if the electron correlation effect is negligible in these compounds.

Another possible contribution is the electron correlation effect, which is not taken into account within the framework of the LDA.
Although the valence band and core-level spectra suggests that the effect is not strong in these compounds, it can have measurable influences to their electronic structures.
There are some studies which include the effect by applying the static mean-field approximations such as the LDA+$U$ \cite{UGe2_LDAU_Shick1, UGe2_LDAU_Shick2, UGe2_LDAU_Yaresko} or the generalized gradient approximation (GGA)+$U$ \cite{UGe2_XPS} for \UG.
These studies could successfully improve the description of the magnetic properties.
Meanwhile, the mean-field type approximation cannot take into account the dynamical nature of \Uf electrons, which changes the electronic structures especially in the vicinity of \EF.
Since the bands are very narrow in these compounds, the renormalization of bands due to the dynamical nature can alter their shapes drastically.
Therefore, the framework such as the LDA + dynamical mean-field theory (DMFT) is needed to account the correlated natures of \Uf electrons in these compounds. 

Recently, Tro\ifmmode \acute{c}\else \'{c}\fi{} \etal suggested that \Uf electrons in \UG consist of the localized and itinerant subsystems \cite{UGe2_dual}.
This argument is based on the fact that the ground state properties of \UG indicate an itinerant character while its high temperature properties are described by the crystal electronic field (CEF) model with the localized $\mathrm{U}~5f^2$ configuration.
If this is the case, most of \Uf contributions should be distributed on the high binding energy side as has been observed in the case of $\mathrm{UPd}_3$ \cite{UPd3_ARPES,UHAXPES}.
However, we have experimentally observed that most of the weight from \Uf states are distributed in the vicinity of \EF, and \Uf electrons form quasi-particle bands even at $T=120~\mathrm{K}$.
Therefore, our results suggest that the high temperature properties of \UG should also be explained by the itinerant \Uf electrons.

The resonant photoemission study of \UCG suggests that \Uf and \Cd states are strongly hybridized in this compound.
This is consistent with the result of the polarized neutron diffraction study where magnetic moments are observed in not only at the U site but also at the $\mathrm{Co}$ site due to the strong $\mathrm{U}~5f-\mathrm{Co}~3d$ hybridization\cite{UCoGe_PND}.
Therefore, the \Cd states play an important role in the ground state properties of this compound.
This is in contrast with the case of \URG, where the $\mathrm{Rh}~4d$ states are distributed on the high-binding-energy side ($E_{\mathrm{B}} = 2-4~\mathrm{eV}$), and they do not have significant contribution to the quasi-particle bands in the vicinity of \EF.

Although the measurement in the ferromagnetic phase of \UCG could not be achieved due to the low transition temperature, it is expected that the shape of Fermi surface in the ferromagnetic phase should be very different from that in the paramagnetic phase since multiple narrow bands are expected in the vicinity of \EF.
For example, the bottom of band 71 along the $\mathrm{Y}-\mathrm{\Gamma}-\mathrm{Y}$ high-symmetry line has its bottom at the $\Gamma$ point at $E_{\mathrm{B}} \sim 5~\mathrm{meV}$, and even very small splitting of bands due to the ferromagnetic transition would result in drastic changes in the topology of Fermi surface.

\section{CONCLUSION}
We have studied the AIPES and ARPES spectra of \UG and \UCG, and compared them with the result of the band structure calculation treating all \Uf electrons as being itinerant.
The results are summarized as follows.

(i) The \Uf electrons in \UG and \UCG form quasi-particle bands in the vicinity of \EF, and they have an itinerant character in the paramagnetic phase.
In particular, the \Uf electrons in \UG have an itinerant character even at $T=120~\mathrm{K}$, suggesting that its physical properties at high temperatures originate from the itinerant \Uf electrons.
Meanwhile, their core-level spectra are accompanied by weak satellite structures on the high-binding-energy side of the main feature, suggesting that electron correlation effects also exist in these compounds.

(ii) The comparison between the ARPES spectra and the result of the band-structure calculations showed that the overall band structures in an energy scale of sub-eV were qualitatively explained by the band structure calculation treating all \Uf electrons as being itinerant.
On the other hand, the states in the vicinity of \EF have more featureless structures than those expected from the calculation, and agreements between the calculation and experiment is not as good as other itinerant \Uf compounds or heavy Fermion uranium compounds.
This means that the Fermi surfaces of \UG and \UCG are qualitatively different from those by the calculation.

(iii) The possible origins of these discrepancies are very complicated band structures of these compounds due to the low symmetry nature of their crystal structures as well as the weak but finite contributions from the electron correlation effect.
To account for the contributions from the electron correlation effect, it is essential to include the dynamical nature of \Uf electrons in the low-symmetry crystals.

(iv) In \UCG, \Uf states are strongly hybridized with \Cd states, suggesting that \Cd states have finite contributions
to electrons at the Fermi level as well as its magnetic properties.

\acknowledgments
The experiment was performed under Proposal Nos 2007A3833 and 2010B3824 at SPring-8 BL23SU.
The present work was financially supported by a Grant-in-Aid for Scientific Research from the Ministry of Education, Culture, Sports, Science, and Technology, Japan, under Contact Nos. 21740271 and 26400374; Grants-in-Aid for Scientific Research on Innovative Areas "Heavy Electrons" (Nos. 20102002 and 20102003) from the Ministry of Education, Culture, Sports, Science, and Technology, Japan; and the Shorei Kenkyu from Hyogo Science and Technology Association.

\appendix
\section{Background subtraction in ARPES spectra}
ARPES spectra of 4$f$ and 5$f$ based compounds are often dominated by strong and sharp peaks at the vicinity of $E_{\mathrm{F}}$\cite{UFeGa5_ARPES, UB2_ARPES, UN_ARPES, UPd2Al3_ARPES1, UPd2Al3_ARPES2}.
This strong signal at $E_{\mathrm{F}}$ has been observed in any part of the momentum space, and this makes it difficult to observe the behavior of quasi-particle bands in the vicinity of $E_{\mathrm{F}}$.

There are some possible origins of this effect.
One possibility is the finite momentum broadenings in ARPES experiments.
There exist the finite broadening effect along $k_\perp$ direction due to the finite escape depth of photoelectrons, as well as finite instrumental momentum resolutions along $k_\parallel$ direction.
Then, the ARPES spectra probe a finite portion of the Brillouin zone, and the peak at $E_{\mathrm{F}}$ might appear in the ARPES spectra even at the point where no peaks at $E_{\mathrm{F}}$ are expected.
This effect is recognized in the simulation of ARPES spectra shown in the right panel of Fig.~\ref{UGe2_ARPES_FS} (a).

Another possibility is the background contributions from elastically scattered photoelectrons.
In the photoemission process, some photoelectrons are scattered by surface disorders or phonons, and they lose their information about momentum.
The latter effect becomes significant when the kinetic energy of photoelectrons increases \cite{Fadley}.
The ARPES spectrum $I(E,\mbox{\boldmath $k$})$ is expressed by the sum of the contributions from direct transition $I_{\rm DT}(E,\mbox{\boldmath $k$})$ and non-direct transition $I_{\rm NDT}(E,\mbox{\boldmath $k$})$\cite{arpes_bg1, arpes_bg2}.

\begin{equation}
I(E,\mbox{\boldmath $k$}) = I_{\rm DT}(E,\mbox{\boldmath $k$}) + I_{\rm NDT}(E,\mbox{\boldmath $k$}),
\end{equation}

$I_{\rm NDT}(E,\mbox{\boldmath $k$})$ depends on the surface morphology in an atomic scale and the Deby-Waller factor.
In particular, the former effect is very difficult to estimate for actual photoemission spectra.
Meanwhile, $I_{\rm NDT}(E,\mbox{\boldmath $k$})$ should have less momentum dependences, and its shape should be similar to that of the angle-integrated spectrum.
Therefore, in the present study, the contribution from the non-direct transition is approximated by the AIPES spectrum $I_{\rm AIPES}(E)$ multiplied by the adjustable parameter $\alpha$

\begin{equation}
I(E,\mbox{\boldmath $k$}) = I_{\rm DT}(E,\mbox{\boldmath $k$}) + \alpha I_{\rm AIPES}(E),
\end{equation}

The parameter $\alpha$ has been chosen to make ARPES spectra reasonable, and typical values were $\alpha=0.6-0.7$.
In the present study, we have used $I_{\rm AIPES}(E)$ as the averaged spectrum along $k_x$ and $k_y$ directions.
By this procedure, the states near $E_{\mathrm{F}}$ become much clearer.

\bibliographystyle{apsrev}
\bibliography{UGe2_UCoGe}

\end{document}